\documentclass[aps,prb,superscriptaddress,twocolumn]{revtex4}
\usepackage{amssymb}
\usepackage{graphicx}
\usepackage{amsmath}
\usepackage[colorlinks=true,linkcolor=blue,citecolor=blue,urlcolor=blue]{hyperref}

\newcommand{\be}{\begin{equation}}
\newcommand{\ee}{\end{equation}}

\newcommand{\bs}{\begin{split}}
\newcommand{\es}{\end{split}}

\usepackage[normalem]{ulem}
\usepackage{amsmath,amssymb}
\usepackage{multirow}
\usepackage{xcolor,soul}
\usepackage{srcltx}
\usepackage{hyperref,graphicx}

\begin{document}

\title{Erratic non-Hermitian skin localization}
\author{Stefano Longhi}
\thanks{stefano.longhi@polimi.it}
\affiliation{Dipartimento di Fisica, Politecnico di Milano, Piazza L. da Vinci 32, I-20133 Milano, Italy}
\affiliation{IFISC (UIB-CSIC), Instituto de Fisica Interdisciplinar y Sistemas Complejos, E-07122 Palma de Mallorca, Spain}

\begin{abstract}
A novel localization phenomenon, termed erratic non-Hermitian skin localization, has been identified in disordered globally-reciprocal non-Hermitian lattices. Unlike conventional non-Hermitian skin effect and Anderson localization, it features macroscopic eigenstate localization at irregular, disorder-dependent positions with sub-exponential decay. Using the Hatano-Nelson model with disordered imaginary gauge fields as a case study, this effect is linked to stochastic interfaces governed by the universal order statistics of random walks. Finite-size scaling analysis confirms the localized nature of the eigenstates. This discovery challenges conventional wave localization paradigms, offering new avenues for understanding and controlling localization phenomena in non-Hermitian physics.
 \end{abstract}
\maketitle

{\em Introduction.} Anderson localization \cite{R1,R2,R3,R4} and the non-Hermitian skin effect (NHSE) \cite{R5,R6,R7,R8,R9,R10,R11,R11b,R11c,R12,R13,R14,R15,R16,R17,R17b,R18,R19,R20,R21,R21b,R21c,R22,R23,R24,R25,R26,R27,R28,R28b,R29,R30,R31} represent two fundamental wave localization phenomena, arising from distinct mechanisms: wave interference in disordered systems and the intrinsic point-gap topology of non-Hermitian (NH) systems, respectively. 
The Hatano-Nelson model has long served as a cornerstone in the study of NH systems, first introduced in seminal works \cite{R32,R33,R34}. Earlier research on this model uncovered a competition between Anderson localization and delocalization induced by an imaginary gauge field \cite{R32,R33,R34,R35,R36,R37,R38,R39,R40,R41,R42,R43}. Recent advances in non-Hermitian and topological physics \cite{R20,R43a,R43b,R43c,R43d,R43e} have unveiled a rich variety of phenomena beyond these initial discoveries, highlighting the exceptional sensitivity of NH systems to boundary conditions, as demonstrated by the NHSE (see, e.g., \cite{R5,R7,R20,R22,R23,R26,R27,R28,R31} and references therein).
This phenomenon, characterized by the accumulation of a large number of eigenstates near the system boundaries \cite{R5}, defies the conventional Bloch band theory and challenges the traditional bulk-boundary correspondence. To reconcile this discrepancy, the generalized Brillouin zone theory \cite{R5,R8,R44}  in one dimensional systems, and amoeba formulation of non-Bloch band theory in higher dimensions \cite{R44bis}, have been introduced. The NHSE can persist in disordered systems with broken translational invariance \cite{R55,R56,R56b,R57,R58,R59,R59b,R59c}.  The interplay between non-Hermiticity and spatial inhomogeneities --such as domain walls, disorder, dislocations or impurities-- gives rise to intriguing localization phenomena \cite{R27}, including 
topological phase transitions \cite{R44a,R44b,R44c,R44d}, impurity-induced topological bound states \cite{R17,R45,R46,R47}, scale-free localization \cite{R48,R48b,R49,R50,R51,R52,R53,R54}, dislocation NHSE \cite{RD1,RD2,RD3}, the inner NHSE \cite{RD4} and unusual form of NH transport \cite{R54a,R54b,R54c,R54d,R54e}. 
\begin{figure}[t]
 \centering
    \includegraphics[width=0.48\textwidth]{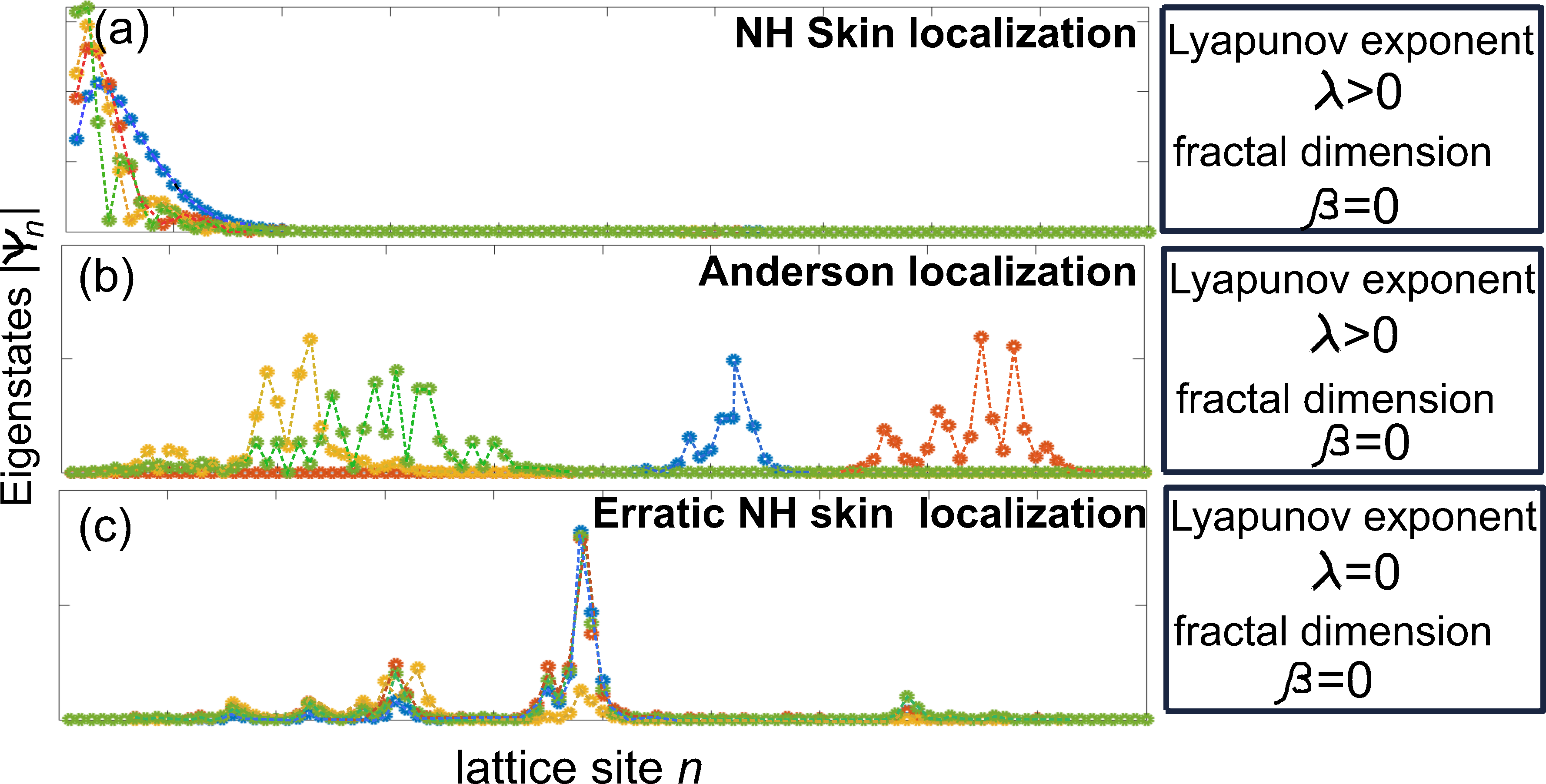}
   \caption{Schematic of three different types of localization in a one-dimensional NH disordered lattice. The three panels depict typical shapes of four eigenstates $|\psi_n|$ in different color.  (a) NH skin localization: all eigenstates are exponentially localized at the lattice edges. (b) Anderson localization: the eigenstates are exponentially localized and their locations are uniformly distributed along the lattice. (c) Erratic skin localization: all eigenstates are localized at around the same position in the lattice, with possible far apart satellite peaks. The localization is lower than exponential and the position of the main and satellite peaks is erratic, i.e. strongly dependent on the realization of disorder.}
    \label{fig1}
\end{figure}

In this Letter, we report a novel form of non-Hermitian localization in one-dimensional disordered and globally-reciprocal lattices, distinct from both Anderson localization and conventional NHSE, which we term the erratic non-Hermitian skin effect (ENHSE). Unlike the NHSE, which involves eigenstate accumulation at specific boundaries [Fig.1(a)], and Anderson localization, where eigenstates are distributed relatively uniformly along the system [Fig.1(b)], the ENHSE is marked by macroscopic eigenstate localization at a seemingly irregular or scattered position throughout the system, with a main localization peak and possible other satellite peaks, depending on the specific realization of disorder [Fig.1(c)]. Unlike Anderson localization or other forms of NHSE in systems with broken translational invariance, such as in the dislocation or inner NHSE  \cite{RD1,RD2,RD3,RD4}, in the ENHSE the localization is {\em lower than exponential}. We illustrate this phenomenon using as a paradigmatic example the Hatano-Nelson model with a spatially fluctuating imaginary gauge field \cite{R55,R56,R57,R59}. The ENHSE is explained in terms of the universal order statistics of random walks \cite{}, where eigenstates macroscopically localize at unpredictable positions in the sample, determined by the disorder realization. Finite-size scaling analysis confirms the localization nature of the eigenstates, which is well-predicted by the statistical properties of gaps in the random walk sequence associated with the fluctuating imaginary gauge field.\\
\par
{\em Hatano-Nelson model with a disordered imaginary gauge field.} To illustrate the phenomenon of ENHSE, let us consider the Hatano-Nelson model \cite{R32} with a spatially-disordered imaginary gauge field \cite{R56,R57,R59}. The model is the described by the NH tight-binding Hamiltonian [Fig.2(a)]
\begin{equation}
\hat{H}=\sum_{n=1}^{N-1} \left( J_n^{R}\hat{c}^{\dag}_{n+1} \hat{c}_n+J_n^L \hat{c}_{n}^{\dag} \hat{c}_{n+1} \right)+ \hat{H}_B
\end{equation}
where $N$ is the number of lattice sites, $\hat {c}_{n}^{\dag}, \hat {c}_{n} $ are the spin-less particle creation  and annihilation
operators at site $n$ ($n=1,2,...,N$), $J^{R,L}_n$ are the right (R) and left (L) hopping amplitudes, and $\hat{H}_B$ is the Hamiltonian term that specifies the lattice boundary conditions.
For open boundary conditions (OBC), one has $\hat{H}_B=0$, whereas for periodic (cyclic) boundary conditions (PBC) one has 
$\hat{H}_B=J_N^{R}\hat{c}^{\dag}_{1} \hat{c}_N+J_N^L\hat{c}_{N}^{\dag} \hat{c}_{1}$. For a  spatially-disordered imaginary gauge field $h_n$, we assume \cite{R59}
\begin{equation}
J_n^R=J \exp(h_n) \; , \;\; J_n^L=J \exp(-h_n).
\end{equation}
where $h_n$ are independent random variables with the same probability density function $f(h)$ of mean value ${\bar h}$ and finite variance $(\Delta h)^2$. The single-particle eigenfunctions $\psi_n$ and corresponding eigenenergies $E$ of $\hat{H}$ satisfy the NH spectral problem
\begin{equation}
E \psi_n=J_n^L \psi_{n+1}+J_{n-1}^R \psi_{n-1}
\end{equation}
with either OBC ($\psi_0=\psi_{N+1}=0$) or PBC ($\psi_{n+N}=\psi_{n}$). As shown in previous works \cite{R56,R57,R59}, the Hamiltonian $\hat{H}$ displays the NHSE provided that the mean value $\bar h$ is non-vanishing.
In fact, after introduction of the non-unitary gauge transformation
\begin{equation}
\psi_n=\phi_n \exp \left( \sum_{l=1}^{n-1} h_l \right)
\end{equation}
the spectral problem (3) reduces to the one of a disorder-free lattice with Hermitian hopping rate $J$, $E \phi_n=J(\phi_{n+1}+\phi_{n-1})$, and with 
$\phi_{0}=\phi_{N+1}=0$ for OBC and $\phi_{N+1}=\phi_{1} \exp(-N \bar{h})$ for PBC, where $\bar{h}=(1/N) \sum_{l=1}^{N} h_l$ is the mean value of $h_n$ in the large $N$ limit. For OBC, the spectral problem is solved by letting $\phi_n= \sin(nq)$ with corresponding real eigenenergy $E_{OBC}=2 J \cos q$, where $q= \alpha \pi/(N+1)$ ($\alpha=1,2,...,N$). For PBC, the spectral problem is solved by letting $\phi_n=\exp(iqn-\bar{h}n)$ with corresponding eigenenergy $E_{PBC}=2 J \cos (q+i \bar h)$, where $q=2 \pi \alpha /N$ ($\alpha=0,1,2,...,N-1$). Hence the energy spectrum is real and described by the interval $(-2J,2J)$ for OBC, wheres it is complex and described by the closed loop (ellipse) $E(q)=2 J \cos (q+i \bar{h})$ for PBC. For $\bar h=0$, the two spectra do coincide [Fig.2(b)], and the NHSE is not anymore observed.\par
{\em Erratic skin localization.} Let us consider the case $\bar{h}=0$. A vanishing value of $\bar{h}$ indicates that the system is globally reciprocal (despite being nonreciprocal at a local level).  In previous work \cite{R59}, it was concluded that in this case the system exhibits characteristics akin to a Hermitian system with all states being delocalized. The absence of both skin and Anderson (exponential) localization of the eigenstates for $\bar{h}=0$, in spite of the disorder in the hopping rates, is indeed confirmed by the vanishing of the Lypaunov exponent $\lambda(E)$ \cite{R56} at any energy $E$ in the spectrum, as shown in Sec.S1 of the Supplemental Material \cite{Supp}.  
\begin{figure}[t]
 \centering
    \includegraphics[width=0.48\textwidth]{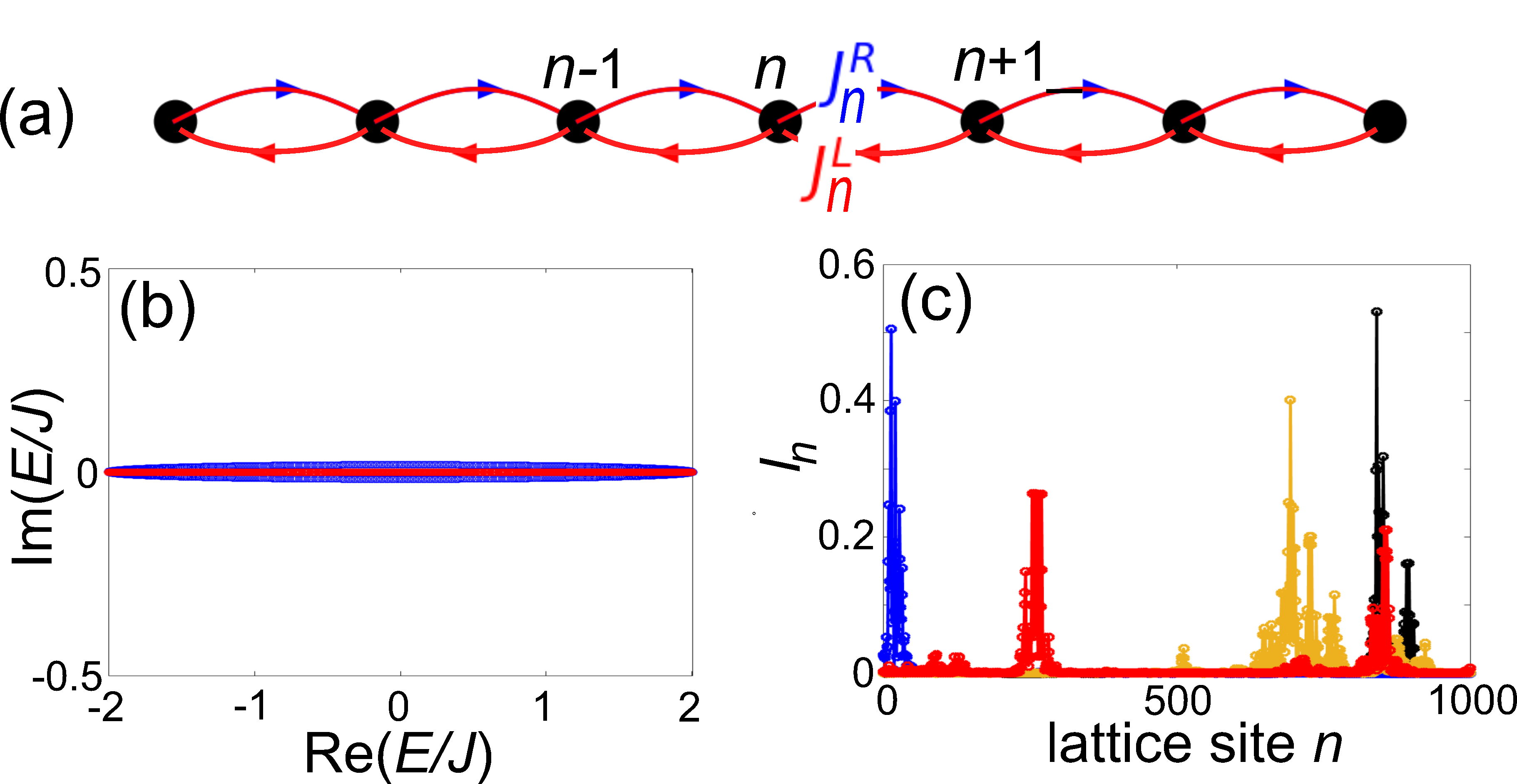}
   \caption{(a) Schematic of the Hatano-Nelson model with disordered imaginary gauge field. (b) Energy spectrum of the Hatano-Neslon model under PBC (blue circles) and OBC (red circles) for a Bernoulli distribution $f(h)$ with $\Delta h=0.4$.  Lattice size $N=1000$. (c) Behavior of the eigenstate distribution $I_n$ under PBC for four different realizations of the stochastic sequence $\{ h_n\}$, indicated by the four different colors. }
    \label{fig2}
\end{figure}
 However, vanishing of the Lypaunov exponent does not necessarily imply an extended state and lack of localization, it just means that the eigenstates are not {\em exponentially} localized. As a matter of fact, an inspection of the shapes of the eigenstates $\psi_n^{(\alpha)}$ ($ \alpha=1,2,3,..,N$) indicates that they are not at all extended. Rather,  all the eigenstates show  a macroscopic localization at some position in the lattice, with typically a main sharp peak and other minor (satellite) peaks. The position of the main and satellite peaks in the lattice is the same for all eigenstates but erratic, i.e. it strongly depends on the specific realization of the sequence $\{ h_n\}$, as shown as an illustrative example in Fig.2(c). The figure depicts the behavior of the spatial eigenstate distribution \cite{R21}, $I_n=(1/N) \sum_{\alpha=1}^N | \psi_n^{(\alpha)}|^2$, for four different realizations of the sequence $\{ h_n \}$. For Anderson localization, $I_n$ would be almost uniform along the sample, whereas for NHSE $I_n$ would be localized at system edges. In the erratic NHSE, $I_n$ displays a main peak with other satellite peaks, erratically located along the sample depending on the disorder realization.
 The localization nature of the eigenstates when $\bar{h}=0$, despite the vanishing of the Lyapunov exponent, is demonstrated by computation of the inverse participation ratio (IPR) and fractal dimension $\beta$, which are commonly used to quantify the localization features of eigenstates in the analysis of Anderson localization \cite{R60,R61,R62,R63,R64,R65}.  For a wave function $\psi_n$, normalized as $ \sum_{n=1}^{N} | \psi_n|^2=1$, the IPR and fractal dimension $\beta$
are defined by ${\rm  IPR}= \sum_{n=1}^{N} | \psi_n |^{4}$ and $\beta= \lim_{N \rightarrow \infty} \frac{\ln {\rm IPR}}{\ln (1/N)}.$ For extended (ergodic) and localized states one has $
   \beta=1$ and $\beta=0$, respectively, whereas $0<\beta<1$ implies multifractality, i.e. critical states. The central result of our analysis is that all eigenstates $\psi_n$ have the same IPR and are localized ($\beta=0$), with a sub-exponential localization as dictated by the vanishing of the Lyapunov exponent. For a fixed value of lattice size $N$, the IPR of the wave functions is a random variable, which depends on the specific realization of the sequence $\{ h_n \}$. A typical probability distribution of IPR is shown in Fig.3(a) for a Bernoulli distribution of $h_n$ ($h_n$ can take with the same probability the two values $ \pm \Delta h$).  The behavior of the mean value of distribution, $\overline{\rm IPR}$, versus lattice size $N$ is shown in Fig.3(b). For comparison, the behavior of the IPR for a  disorder-free lattice, with $h_n=0$, is also shown.  
   \begin{figure}[t]
 \centering
    \includegraphics[width=0.48\textwidth]{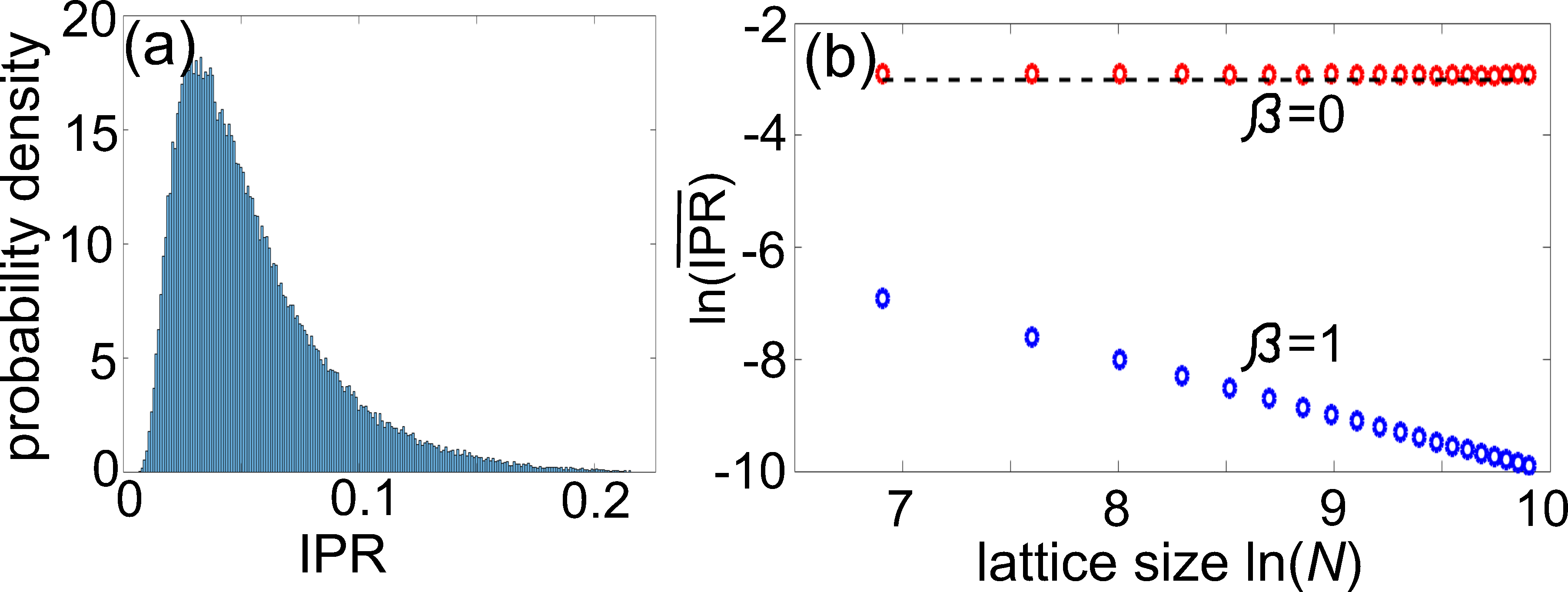}
   \caption{(a) Probability density function of the IPR in a lattice of size $N=2000$ with PBC and a disordered gauge field with a Bernoulli distribution ($\Delta h=0.4$). The probability distribution has been numerically computed by considering $10^5$ realizations of the sequence $\{ h_n \}$. (c) Behavior of the mean value $\overline{\rm{IPR}}$ of the IPR distribution versus lattice size (red circles) on a log scale. The dashed curve is the theoretical prediction based on Eq.(5). The blue circles show, for comparison, the IPR behavior in a disorder-free lattice ($h_n=0$). $\beta$ is the corresponding fractal dimension.}
    \label{fig3}
\end{figure}
       The figure clearly indicates that $\overline{\rm{ IPR}}$  does not vanish as $N$ is increased, and reaches a stationary value, corresponding to a fractal dimension $\beta=0$ and localized eigenstates.
       Similar results are obtained for other probability density functions $f(h)$ of the fluctuating gauge field $h_n$, {\color{black} such as for uniform and normal distributions  (see Fig.S1 of the Supplemental Material \cite{Supp})}.\\
        The physical origin of such a novel kind of localization is very distinct than both Anderson and skin localization, and is rooted in the extreme value statistics of random walks \cite{R66,R67,R68}. In fact, as shown in Sec.S2 of the Supplemental Material \cite{Supp} under PBC the IPR is the same for all eigenfunctions, i.e. it does not depend on the energy $E$, and can be evaluated in terms of the statistical properties of the stochastic process $X_n=\sum_{l=1}^{n-1} h_l$ with $X_1=0$. This process effectively describes a discrete-time symmetric random walk on a line, where at each time step $n$ the walker with a probability density function $f(h)$ erratically shifts by a quantity $h$, either positive or negative, on the line, as schematically shown in Fig.4(a). Clearly, according to Eq.(4) the main and satellite peaks of the wave functions correspond to the extreme positive positions of the walker (absolute and relative maxima of $X_n$) during the random walk, which depend on the specific walk realization. Basically, around each local maximum of $X_n$ a local interface with opposite signs of the imaginary gauge field is realized, toward which the excitation is pushed via a NHSE at the interface \cite{R17,R46,R47}. Hence the main and satellite peaks of the wave functions correspond to local skin localization at the gauge field interfaces erratically created by the random walk $X_n$ [Fig.4(b)]. This local interface picture is not enough to demonstrate the global localization nature of the wave functions.
 To prove localization, let us rearrange the random variables $X_n$ in decreasing order of magnitude $M_{1}\geq M_2 \geq ...>M_k\geq ... \geq M_n$, where $M_k$ is the $k$-th maximum of the set $\{ X_1,X_2,...,X_n \}$, with $M_1={\rm max}_k X_k$ and $M_n={\rm min}_k X_k$. Indicating by $n_k$ the lattice position of the $k$-th maximum $M_k$, i.e. such that $X_{n_k}=M_k$, the wave function $\psi_n$ is characterized by a sequence of decreasing peaks, the main one located at the site $n=n_1$. The IPR of the wave function depends on the statistical distribution of the {\em gaps} $\delta_k=M_{k}-M_{k+1}$ between successive maxima [Fig.4(c)], which in the large $n$ limit displays a universal behavior rooted in the extreme value statistics of random walks \cite{R66}. As shown in Sec.S2 of the Supplemental Material \cite{Supp}, 
an approximate expression of the mean value $\overline{ \rm{IPR}}$ of the IPR distribution can be derived as
\begin{figure}[t]
 \centering
    \includegraphics[width=0.48\textwidth]{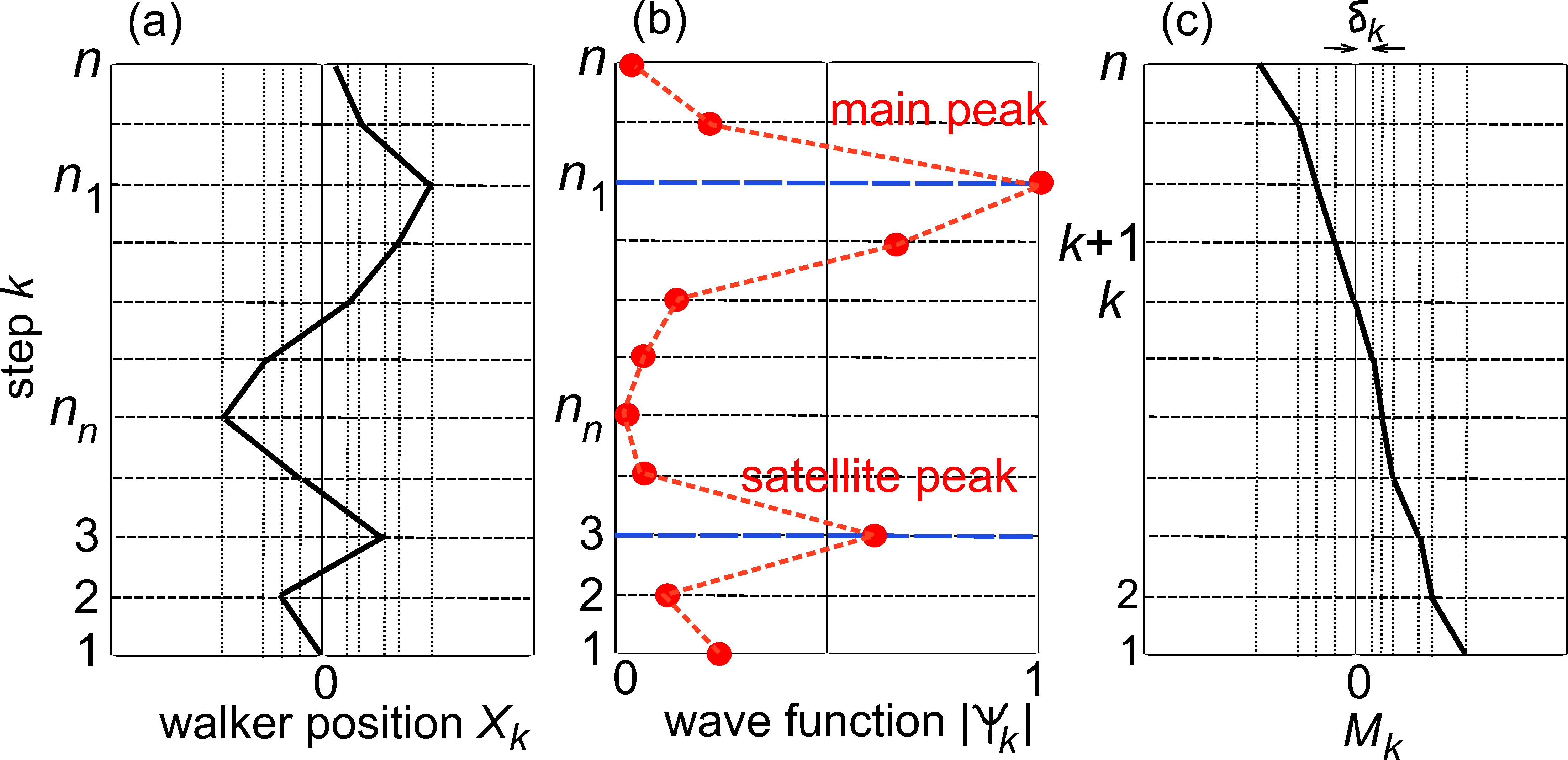}
   \caption{(a) Schematic of the random walk $X_n=\sum_{k=1}^{n-1} h_l$ defined by the stochastic sequence $\{ h_k \} $ of the  gauge field. Note that for the specific walk realization there are two relative maxima of $X_k$, at $k=n_1$ and $k=3$, which define two interfaces around which the gauge field $h$ locally displays opposite signs. (b) Schematic of the eigenfunction $|\psi_k|$, displaying a main peak and a satellite peak at the two interfaces. (c) Ordering procedure used to compute the IPR; $\delta_k$ are the gaps between adjacent ordered values of $X_k$.}
    \label{fig4}
\end{figure}
\begin{equation}
{\rm \overline{IPR}} \simeq \frac{1+\sum_{k=2}^{N} \exp(-4Y_k) }{\left(1+ \sum_{k=2}^{N} \exp(-2Y_k) \right)^2} 
\end{equation}
where we have set $Y_k= \Delta h  \sum_{l=1}^{k-1} \sqrt{1/ (2 \pi l)}$. For large $k$, one has $Y_k \simeq \Delta h  \sqrt{ 2 k/ \pi}$, and hence for $N \rightarrow \infty$ the series appearing in the numerator and denominator of Eq.(5) converge to finite values. This means that, as $N \rightarrow \infty$, the $ {\rm \overline{IPR}}$ reaches a stationary and non-vanishing value, dependent solely on the variance $\Delta h$ of the fluctuating gauge field $h_n$, indicating that the wave functions are localized. The theoretically predicted value of $\overline{ \rm{IPR}}$ versus $N$, given Eq.(5), fits very well with the numerical results, as shown in Fig.3(b).\par
\begin{figure}
 \centering
    \includegraphics[width=0.48\textwidth]{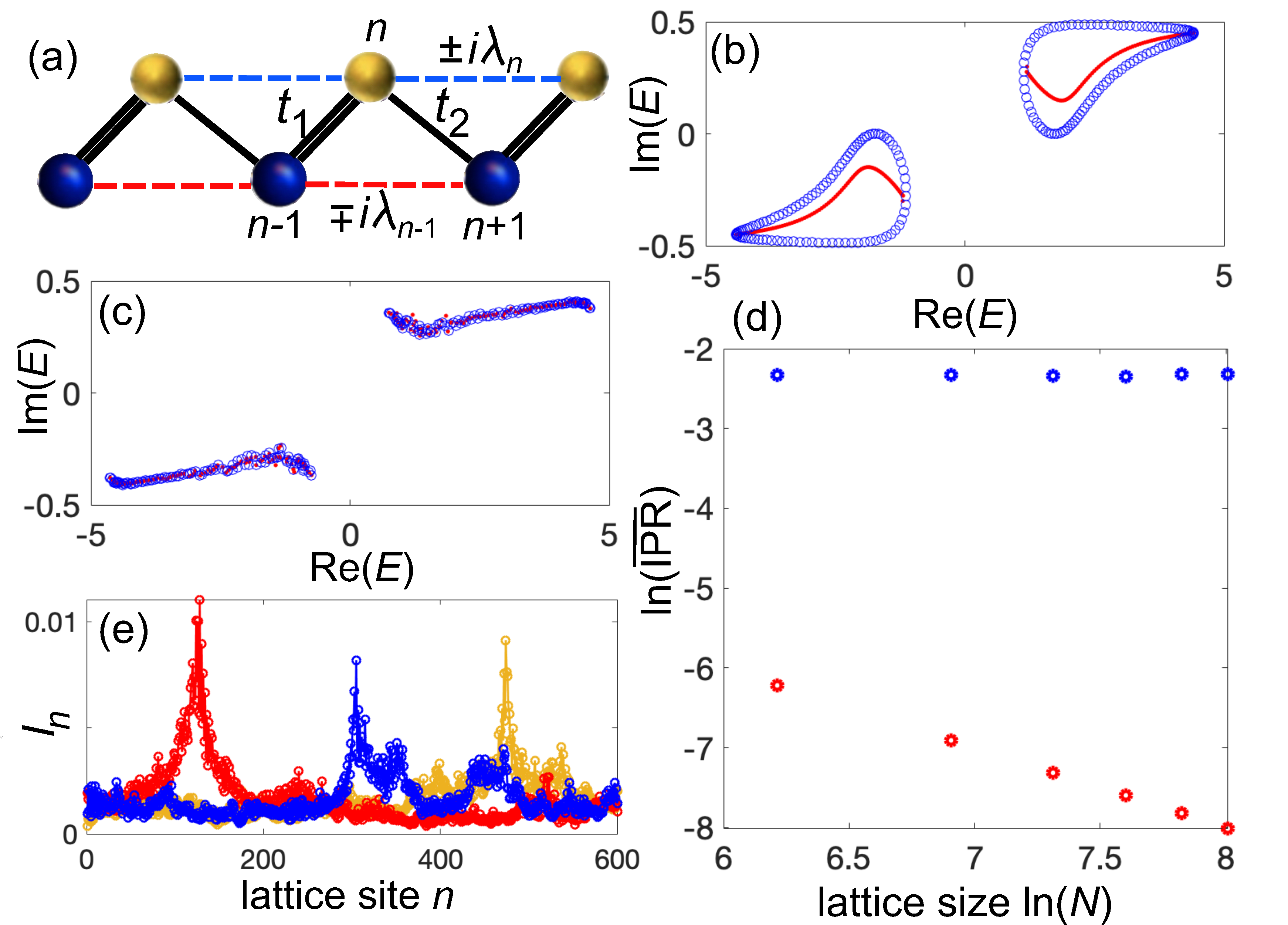}
   \caption{(a) Schematic of the NH Rice-Mele model displaying the reciprocal NHSE. (b) Energy spectrum under PBC (blue circles) and OBC (red points) in the disorder-free lattice for $t_1=1$, $t_2=1.5$, $\Delta=2$, $\gamma=0.5$ and $\lambda=1$. (c) Same as (b), but in the disordered lattice, where each amplitude in the sequence $\{ \lambda_n  \}$ can take only the two values $\pm 1$ with equal probability (Bernoulli distribution). (d) Behavior of the mean IPR versus system size $N$ on a log scale (blue circles); a statistical average over 100 different realizations of the sequence $\{ \lambda_n \}$ has been assumed. For comparison, the red circles show the behavior IPR$=1/N$ of a delocalized phase. (e) Behavior of the eigenstate distribution $I_n$ for three different realizations of the stochastic sequence $\{ \lambda_n\}$ (lattice size $N=600$). }
    \label{fig5}
\end{figure}
The erratic skin localization effect discussed above is expected to be a general phenomenon in one-dimensional (or quasi-one dimensional) lattices which display local non-reciprocity but that are globally reciprocal. As an additional example, in Sec.S3 of the Supplemental Material \cite{Supp} the erratic skin localization is shown to arise in two side-coupled Hatano-Nelson chains, which in the disorder-free regime displays the critical NHSE \cite{R14,R48b}. 
It is also important to note that erratic skin localization is robust against small but finite values of the disorder bias, i.e., when the average $\bar{h} \neq 0$. In this regime, the associated stochastic process $X_n$ describes a biased discrete-time random walk, and the energy spectrum as well as the localization properties of the eigenstates become sensitive to boundary conditions. Under PBC, the energy spectrum exhibits a nontrivial point-gap topology, while the eigenstate localization remains qualitatively similar to the case of zero bias ($\bar{h} = 0$). This implies that erratic non-Hermitian localization persists under PBC, even in a globally non-reciprocal lattice. In contrast, under OBC, the situation changes due to non-Hermitian pumping toward the system's edge. Specifically, when the bias $\bar{h}$ is much smaller than the standard deviation $\Delta h$ of the disorder distribution $f(h)$, the erratic formation of skin interfaces in the bulk dominates, leading to localization away from the edge. However, as the bias $\bar{h}$ increases and becomes comparable to $\Delta h$, edge localization progressively takes over, and the conventional NHSE is gradually restored.

{\it Erratic skin localization in a reciprocal model.} In one-dimensional lattices, the NHSE can arise in reciprocal systems as well, i.e. in the absence of imaginary gauge fields  (see e.g. \cite{R17b,R27,R69,R70}). It is thus worth considering the emergence of erratic skin localization in reciprocal disordered models. As an illustrative example, let us consider the two-band NH Rice-Mele model \cite{R17b} with local dissipation/gain $\gamma$ [Fig.5(a)], where stochastic disorder is assumed for the hopping amplitudes $\lambda_l$ in the same sublattices. The Hamiltonian of the system reads \cite{R17b,R69}
 \begin{eqnarray}
\hat{H} & = & \sum_{n=1}^{N-1} \left( J_n\hat{c}^{\dag}_{n+1} \hat{c}_n+ {\rm H.c.} \right)-\sum_{n=1}^N (-1)^n (\Delta+i \gamma)  \hat{c}^{\dag}_{n} \hat{c}_n \nonumber \\
&+ &  \sum_{n=1}^{N-2} (-1)^n \left( i \lambda_n \hat{c}^{\dag}_{n+2} \hat{c}_n+ {\rm H.c.} \right) + \hat{H}_B
\end{eqnarray}
where $N$ is the (even) number of lattice sites, $J_n=t_1$ for $n$ odd and $J_n=t_2$ for $n$ even, $\Delta$ and $\gamma$ are local energy shift and gain/loss terms, and the hopping $\lambda_n$ are independent random variables with the same probability density function $f(\lambda)$ of zero mean. $\hat{H}_B$ is the Hamiltonian term that specifies the lattice boundary condition, either OBC or PBC.  In the disorder-free model, where $\lambda_n=\lambda$ is homogeneous across the lattice, the PBC and OBC energy spectra for the two bands differ, signaling the presence of the NHSE \cite{R17b,R69} [Fig.5(b)]. This suggests that, even though the system does not exhibit non-reciprocal hopping, projecting onto a specific subspace --such as in one of the two sublattices-- can reveal effective non-reciprocal hopping within that subspace. In the disordered model, the OBC and PBC spectra are almost identical in the large $N$ limit [Fig. 5(c)], indicating that the NHSE is no longer present. Instead, we observe a form of erratic skin localization similar to the single-band Hatano-Nelson model. This behavior is clearly demonstrated in Figs. 5(d) and 5(e). For a given realization of the sequence $\{ \lambda_n \}$, we calculate the global IPR of the eigenstates $\psi_n^{(\alpha)}$ under PBC, i.e. ${\rm IPR}=(1/N) \sum_{n, \alpha=1}^N |\psi_n^{(\alpha)}|^4$, and then a statistical average, $\overline{\rm{IPR}}$, is made for different stochastic realizations of the sequence $\{ \lambda_n \}$. As one can see, the ${\rm \overline{IPR}}$ does not decreases as the system size $N$ increases, indicating the localization of eigenstates on average. A typical behavior of  eigenstate distribution $I_n$ is shown Fig.5(e), clearly indicating the erratic (i.e. disorder-dependent) nature of localization.\\
{\color{black}Although the preceding analysis
focuses on systems with non-Hermitian Hamiltonians, as shown in Sec.4 of the Supplemental Material \cite{Supp} a similar dynamics can arise in open quantum systems described by a Lindblad master equation, where effective non-reciprocal hopping arises from the interplay of coherent and dissipative couplings \cite{R56b}.}\par
 {\em Conclusion.}
A novel localization phenomenon, termed erratic non-Hermitian skin localization, has been identified in disordered non-Hermitian lattices. Unlike the conventional non-Hermitian skin effect and Anderson localization, this phenomenon exhibits macroscopic eigenstate localization at irregular, disorder-dependent positions, characterized by sub-exponential decay.
Using the Hatano-Nelson model with disordered imaginary gauge fields as a case study of globally reciprocal non-Hermitian lattices, this effect is attributed to stochastic interfaces governed by the universal order statistics of random walks. Lyapunov exponent and finite-size scaling analysis confirm the sub-exponential localized nature of the eigenstates.
Erratic non-Hermitian skin localization challenges conventional understandings of wave localization in non-Hermitian systems, offering fresh insights into disorder-induced phenomena and unlocking new possibilities for wave manipulation and control in engineered systems. It also invites further investigation and potential extensions to higher-dimensional NH systems, where different types of skin localization can arise\cite{R21,R21b,R23,R27}.\\
\\
  {\em Acknowledgments.}
The author acknowledges the Spanish State Research Agency, through the Severo Ochoa
and Maria de Maeztu Program for Centers and Units of Excellence in R\&D (Grant No. MDM-2017-0711).

\end{document}